# Measurement of Gilbert damping parameters in nanoscale CPP-GMR spin-valves

Neil Smith, Matthew J. Carey, and Jeffrey R. Childress.
San Jose Research Center
Hitachi Global Storage Technologies
San Jose, CA 95120

*abstract*—In-situ, device level measurement of thermal mag-noise spectral linewidths in 60nm diameter CPP-GMR spin-valve stacks of IrMn/ref/Cu/free , with reference and free layer of similar CoFe/CoFeGe alloy, are used to simultaneously determine the intrinsic Gilbert damping for both magnetic layers. It is shown that careful alignment at a "magic-angle" between free and reference layer static equilibrium magnetization can allow direct measurement of the broadband intrinsic thermal spectra in the virtual absence of spin-torque effects which otherwise grossly distort the spectral line shapes and require linewidth extrapolations to zero current (which are nonetheless also shown to agree well with the direct method). The experimental magic-angle spectra are shown to be in good qualitative and quantitative agreement with both macrospin calculations and micromagnetic eigenmode analysis. Despite similar composition and thickness, it is repeatedly found that the IrMn exchange pinned reference layer *has ten times larger* intrinsic Gilbert damping $(\alpha \approx 0.1)$ than that of the free-layer $(\alpha \approx 0.01)$ .It is argued that the large reference layer damping results from strong, off -resonant coupling to to lossy modes of an IrMn/ref couple, rather than commonly invoked two-magnon processes.

## I. INTRODUCTION

Spin-torque phenomena, in tunneling magnetoresistive (TMR) or giant-magnetoresistive (GMR) film stacks lithographically patterned into ~100 nm nanopillars and driven with dc electrical currents perpendicular to the plane (CPP) of the films have in recent years been the topic of numerous theoretical and experimental papers, both for their novel physics as well as potential applications for magnetic memory elements, microwave oscillators, and magnetic field sensors and/or magnetic recording heads.[1] In all cases, the electrical current density at which spin-torque instability or oscillation occurs in the constituent magnetic film layers is closely related to the magnetic damping of these ferromagnetic (FM) films

This paper considers the electrical measurement of thermal mag-noise spectra to determine intrinsic damping at the *device level* in CPP-GMR spin-valve stacks of sub-100nm dimensions (intended for read head applications), which allows simultaneous *R-H* and transport characterization on the same device. Compared to traditional ferromagnetic resonance (FMR) linewidth measurements at the bulk film level, the device-level approach naturally includes finite-size and spin-pumping[2] effects characteristic of actual devices, as well as provide immunity to inhomogeneous and/or two-magnon linewidth broadening not relevant to nanoscale devices. Complimentary to spin-torque-FMR using *ac* excitation currents,[3] broadband thermal excitation naturally excites all modes of the system (with larger, more quantitatively modeled signal amplitudes) and allows simultaneous damping measurement in both reference and free FM layers of the spin-valve, which will be shown to lead to some new and unexpected conclusions. However, spin-torques at finite dc currents can substantially alter the absolute linewidth, and so it is necessary to account for or eliminate this effect in order to determine the intrinsic damping.

## II. PRELIMINARIES AND MAGIC-ANGLES

Fig. 1a illustrates the basic film stack structure of a prospective CPP-GMR spin-valve (SV) read sensor, which apart from the Cu spacer between free-layer (FL) and reference layer (RL), is identical in form to well-known, present day TMR sensors. In addition to the unidirectional exchange coupling between the IrMn and the pinned-layer (PL), the usual "synthetic-antiferromagnet" (SAF) structure PL/Ru/RL is meant to increase magnetostatic stability and immunity to field-induced rotation of the PL-RL couple, as well as strongly reduce its net demagnetizing field on the FL which otherwise can rotate in response to signal fields. However, for simplicity in interpreting and modeling the spectral and transport data of Sec. III , the present experiment restricts attention to devices with a single RL directly exchange-coupled to IrMn, as shown in Fig. 1b.

The simplest practical model for describing the physics of the device of Fig. 1b is a macrospin model that treats the RL unit magnetization $\hat{\boldsymbol{m}}_{\rm RL}$ as fixed, with only the FL magnetization $\hat{\boldsymbol{m}}_{\rm FL}(t) \leftrightarrow \hat{\boldsymbol{m}}(t)$ as possibly dynamic in time. As was described previously,[4] the *linearized* Gilbert equations for small deviations $\boldsymbol{m}' = (m'_{y'}, m'_{z'})$ about equilibrium $\hat{\boldsymbol{m}}_0 \leftrightarrow \hat{\boldsymbol{x}}'$ can be expressed in the primed coordinates as a 2D tensor/matrix equation[5]:

$$(\overleftrightarrow{D} + \overleftrightarrow{G}) \cdot \frac{d\boldsymbol{m}'}{dt} + \overleftrightarrow{H}' \cdot \boldsymbol{m}' = \boldsymbol{h}'(t) \equiv p\, \frac{\partial \boldsymbol{m}'}{\partial \hat{\boldsymbol{m}}} \cdot \boldsymbol{h}(t)$$

$$\overleftrightarrow{D} \equiv \alpha \frac{p}{\gamma} \begin{pmatrix} 1 & 0 \\ 0 & 1 \end{pmatrix}, \; \overleftrightarrow{G} \equiv \frac{p}{\gamma} \begin{pmatrix} 0 & -1 \\ 1 & 0 \end{pmatrix}, \; p \equiv \frac{(M_s V)_{\rm FL}}{\Delta m} \qquad (1)$$

$$\overleftrightarrow{H}' \equiv (\boldsymbol{H}^{\rm eff} \cdot \hat{\boldsymbol{m}}_0) \begin{pmatrix} 1 & 0 \\ 0 & 1 \end{pmatrix} - \frac{\partial \boldsymbol{m}'}{\partial \hat{\boldsymbol{m}}} \cdot \frac{\partial \boldsymbol{H}^{\rm eff}}{\partial \hat{\boldsymbol{m}}} \cdot \frac{\partial \hat{\boldsymbol{m}}}{\partial \boldsymbol{m}'}$$

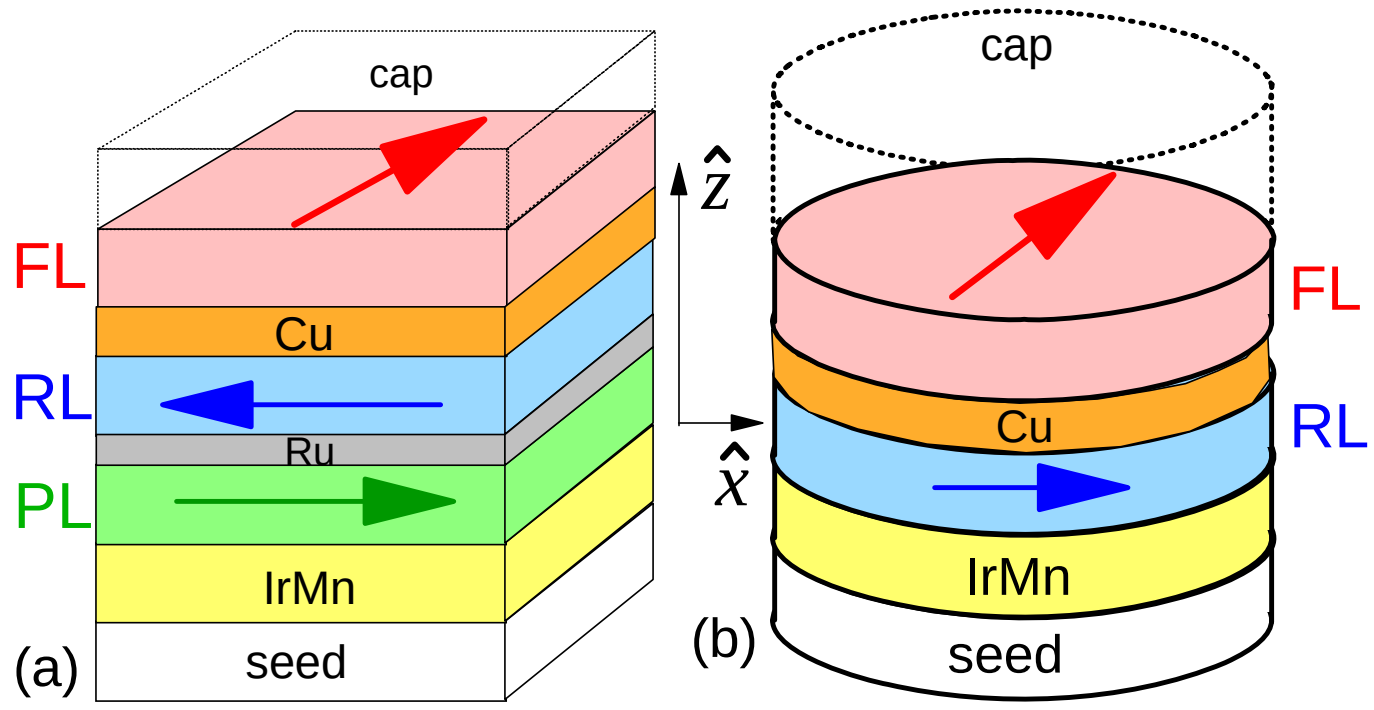

FIG. 1. (a) Cartoon of prospective CPP-GMR spin-valve sensor stack, analogous to that used for contemporary TMR read head. (b) Cartoon of simplified spin-valve stack used for present experiments, patterned into ~60nm circular pillars using e-beam lithography.

In (1), $\partial \boldsymbol{H}^{\text{eff}} / \partial \hat{\boldsymbol{m}}$ is a 3D Cartesian tensor, $\partial \hat{\boldsymbol{m}} / \partial \boldsymbol{m}'$ is a $3\times 2$ transformation matrix between 3D unprimed and 2D primed vectors (with $\partial \boldsymbol{m}' / \partial \hat{\boldsymbol{m}}$ its transpose) which depends only on $\hat{\boldsymbol{m}}_0$, and $\boldsymbol{h}(t)$ is a 3D perturbation field supposed as the origin of the deviations $\boldsymbol{m}'(t)$. The magnetic moment $\Delta m$ is an arbitrary *fixed* value, but $\Delta m \to (M_s V)_{\text{FL}}$ is a natural choice for Sec. II. Using an explicit Slonczewski[6] type expression for the spin-torque contribution, the general form for $\boldsymbol{H}^{\text{eff}}(\hat{\boldsymbol{m}})$ is

$$\boldsymbol{H}^{\text{eff}} = \frac{-1}{\Delta m}\frac{\partial E}{\partial \hat{\boldsymbol{m}}} - \eta(\cos\theta)\, H_{\text{ST}}\, (\hat{\boldsymbol{m}}_{\text{RL}} \times \hat{\boldsymbol{m}}_{\text{FL}}),$$
$$\cos\theta \equiv \hat{\boldsymbol{m}}_{\text{RL}} \cdot \hat{\boldsymbol{m}}_{\text{FL}}, \text{ and } H_{\text{ST}} \equiv (\hbar / 2e)\, P_{\text{eff}}\, J_e / \Delta m \qquad (2)$$

for any free energy function $E(\hat{\boldsymbol{m}})$. A positive electron current density $J_e$ implies electron flow from the RL to the FL. $P_{\text{eff}}$ is the net spin polarization of the current inside the Cu spacer. Oersted-field contributions to $\boldsymbol{H}^{\text{eff}}$ will be neglected here.

With $\boldsymbol{h}(t) = 0$ in (1), nontrivial solutions $\boldsymbol{m}'(t) = \boldsymbol{m}' e^{-st}$ require $s$ satisfy $\det|\overleftrightarrow{H} - s(\overleftrightarrow{D} + \overleftrightarrow{G})| = 0$. The value $J_e \equiv J_e^{\text{crit}}$ when $\operatorname{Re} s = 0$ defines the critical onset of spin-torque instability. Using (1), the general criticality condition is expressible as

$$\underbrace{\alpha\,(H'_{y'y'} + H'_{z'z'})}_{J_e\text{-independent}} + \underbrace{H'_{y'z'} - H'_{z'y'}}_{\propto J_e} = 0 \qquad (3a)$$

$$\frac{H'_{y'z'} - H'_{z'y'}}{H_{\text{ST}}} \cong (1-q^2)\frac{d\eta}{dq} - 2q\,\eta(q) \quad (q \equiv \cos\theta) \qquad (3b)$$

$$\Rightarrow J_e^{\text{crit}} = \frac{\Delta m}{(\hbar/2e)P_{\text{eff}}}\, \frac{\alpha\,(H'_{y'y'} + H'_{z'z'})}{(1-q^2)\,d\eta/dq - 2q\,\eta(q)} \qquad (3c)$$

where $\alpha$ is the Gilbert damping. The $J_e$-scaling of the terms in in (3a) follows just from the form of (2). The result in (3b) was derived earlier[4] in the present approximation of rigid $\hat{\boldsymbol{m}}_{\text{RL}}$.

With $\theta$ the angle between $\hat{\boldsymbol{m}}_{\text{RL}}$ and $\hat{\boldsymbol{m}}_{\text{FL}}$ (at equilibrium), it follows from (3c) that at a "magic-angle" $\theta_{\text{magic}}$ where the denominator vanishes, $J_e^{\text{crit}} \to \infty$ *and spin-torque effects are effectively eliminated from the system at finite* $J_e$. To pursue this point further, explicit results for $\eta(\cos\theta)$ will be used from the prototypical case where the CPP-GMR stack (Fig. 1b) is approximately symmetric about the Cu spacer, which is roughly equivalent to the less restrictive situation where the RL and FL are similar materials with thicknesses that are *not* small compared the spin-diffusion length. For this quasi-symmetric case, both quasi-ballistic[6] and fully diffusive[7] transport models yield the following simple functional forms:

$$\eta(\cos\theta) = \Gamma / [\Gamma + 1 + (\Gamma - 1)\cos\theta]$$
$$r(\cos\theta) \equiv \frac{\delta R \equiv R(\cos\theta) - R_{\text{min}}}{\Delta R \equiv R_{\text{max}} - R_{\text{min}}} = \frac{\eta(\cos\theta)}{\Gamma}(1 - \cos\theta) \qquad (4)$$

which also relates $\eta$ to the normalized resistance $r$ $(0 \le r \le 1)$ which is directly measurable experimentally. The transport parameter $\Gamma$ is theoretically related to the Sharvin resistance[6,8] or mixing conductance[8] at the Cu/FL interface, but will be estimated via $J_e^{\text{crit}}$ measurement in Sec III. Using $\eta(q)$ from (4) in (3), $\theta_{\text{magic}}$ and $r(\theta_{\text{magic}})$ vs. $\Gamma$ curves are shown in Fig. 2.

The "magic-angle" concept also applies to mag-noise power spectral density (PSD) $S_V = [I_{\text{bias}} \Delta R\, (dr/d\theta)_{\text{bias}}]^2 S_\theta$ at bias current $I_{\text{bias}}$, arising from thermal fluctuations in $\theta$ about equilibrium bias angle $\theta_{\text{bias}}$. Assuming $\hat{\boldsymbol{m}}_{0\text{-RL,FL}}$ in/near the film plane ($\hat{\boldsymbol{z}}' \cong \hat{\boldsymbol{z}}$ = plane-normal), and *requiring* $|I_{\text{bias}}| < |I_e^{\text{crit}}|$, it can be shown[5] from fluctuation-dissipation arguments that

$$S_\theta(f) \cong \frac{4k_B T}{\gamma \Delta m}(\overleftrightarrow{\chi} \cdot \overleftrightarrow{D} \cdot \overleftrightarrow{\chi}^{\dagger})_{y'y'},\ \overleftrightarrow{\chi}(\omega) = [\overleftrightarrow{H}' - i\omega(\overleftrightarrow{D} + \overleftrightarrow{G})]^{-1}$$

$$\Rightarrow S_\theta(f) \cong \frac{4\gamma k_B T \alpha}{\Delta m}\, \frac{\gamma^2 (H'^2_{z'z'} + H'^2_{y'z'}) + \omega^2}{(\omega^2 - \omega_0^2)^2 + (\omega\Delta\omega)^2} \qquad (5)$$

$$\text{where } \omega_0 = \gamma\sqrt{H'_{y'y'}H'_{z'z'} - H'_{y'z'}H'_{z'y'}}$$
$$\text{and } \Delta\omega = \gamma[\,\alpha\,(H'_{z'z'} + H'_{y'y'}) + H'_{z'y'} - H'_{y'z'}\,]$$

Comparing (5) with (3), it is seen that the spectral linewidth $\Delta\omega$ is predicted to be a linear function of $J_e$, *but with* $d\Delta\omega / dJ_e \to 0$ *when* $\theta_{\text{bias}} \to \theta_{\text{magic}}$. Since $H'_{z'z'} \gg H'_{y'y'}$ (due to ~10 kOe out-of-plane demag fields) and $H'_{y'y'} \gg H'_{y'z'}$ (e.g., for the $J_e < J_e^{\text{crit}}$ measurements in Sec. III), it is only in the linewidth $\Delta\omega$ that the off-diagonal terms $H'_{y'z'}, H'_{z'y'}$ can be expected to influence $S_\theta(f)$. Therefore, measurement of $S_V(f)$ with $\theta_{\text{bias}} \cong \theta_{\text{magic}}$ ideally allows direct measurement of the natural thermal-equilibrium mag-noise spectrum $S_\theta(f)$, from which can be extracted the intrinsic (i.e., $J_e$-independent) Gilbert damping constant $\alpha$. This is the subject of Sec. III.

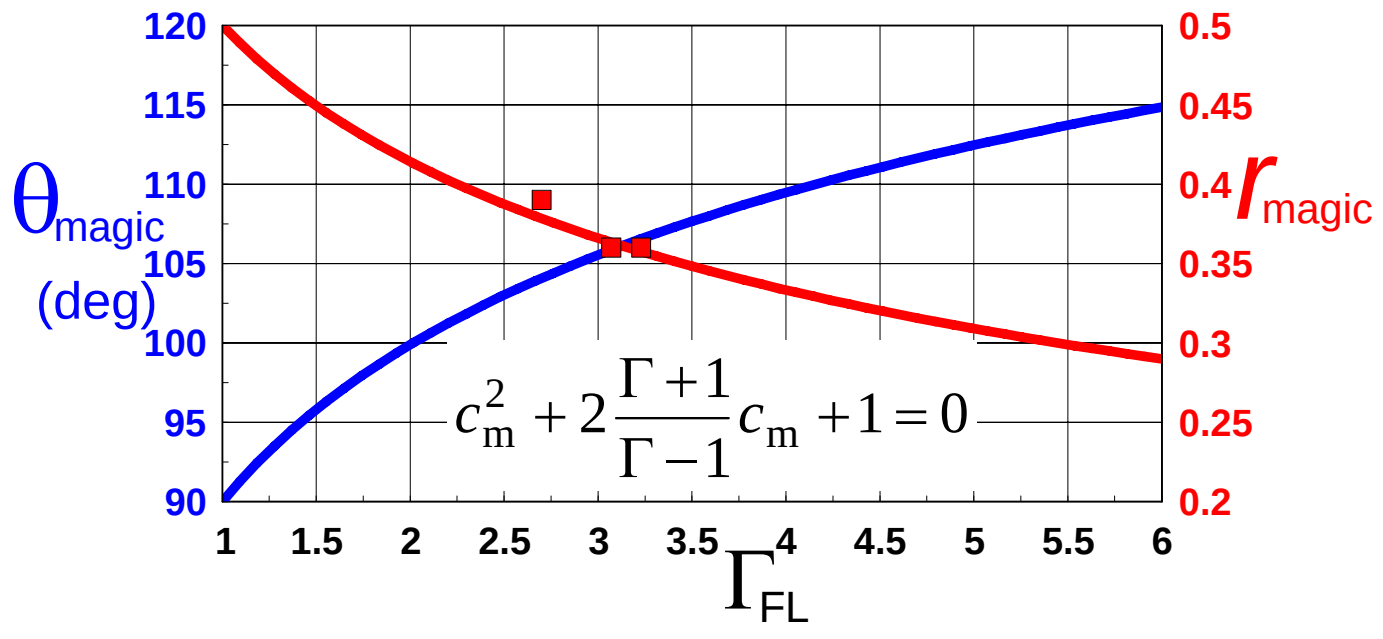

FIG. 2. Graph of $\theta_{\text{magic}}$ (blue) and $r_{\text{magic}} = r_{\text{bias}}(\theta_{\text{magic}})$ (red) vs. $\Gamma_{\text{FL}}$ as described by (4). The equation for $c_{\text{m}} = \cos(\theta_{\text{magic}})$ follows from (3) and (4). The red solid squares are measured $(\Gamma_{\text{FL}}, r_{\text{magic}})$ from Figs. 3,4 and 6.

## III. EXPERIMENTAL RESULTS

The results to be shown below were measured on CPP-GMR-spin-valves of stack structure: seed-layers/IrMn (60A)/RL/Cu (30A)/FL/cap layers. The films were fabricated by magnetron sputtering onto AlTiC substrates at room temperature, with 2mTorr of Ar sputter gas. The bottom contact was a ~1-µm thick NiFe layer, planarized using chemical-mechanical polishing. To increase $\Delta R/R$, both the RL and FL were made from $(CoFe)_{70}Ge_{30}$ magnetic alloys.[9] The RL includes a thin CoFe between IrMn and CoFeGe to help maximize the exchange coupling strength, and both RL and FL include very thin CoFe at the Cu interface. The resultant $M_s t$ product for the RL and FL were about 0.64 emu/cm$^2$. After deposition, SV films were annealed for 5hours at 245C in 13kOe applied field to set the exchange pinning direction. The IrMn/RL exchange pinning strength of ≈0.75 erg/cm$^2$ was measured by vibrating sample magnetometry. After annealing, patterned devices with ≈ 60 nm diameter (measured at the FL) were fabricated using e-beam lithography and Ar ion milling. A 0.2µm-thick Au layer was used as the top contact to devices.

Fig. 3 illustrates a full measurement sequence. Devices are first pre-screened to find samples with approximate ideal in-plane $\delta R$-$H$ loops (Fig. 3a) for circular pillars: non-hysteretic, unidirectionally-square loops with $H = H_{\parallel}$ parallel with the RL's exchange pinning direction ($+\hat{\boldsymbol{x}}$), along with symmetric loops about $H = 0$ when $H = H_{\perp}$ is transverse ($\hat{\boldsymbol{y}}$-axis). The right-shift in the $\delta R$-$H_{\parallel}$ loop indicates a large demagnetizing field of ~500 Oe from the RL on the FL.

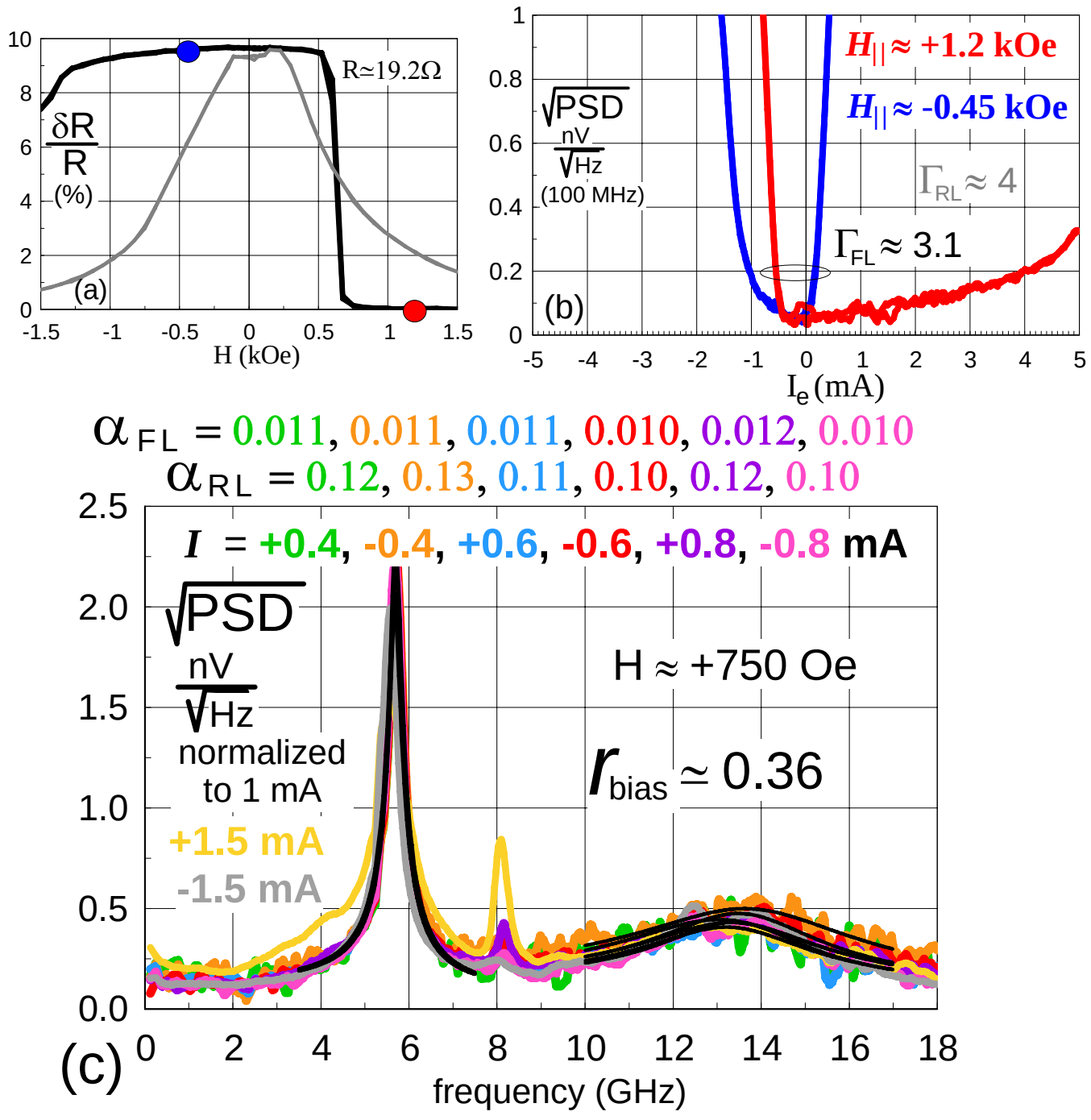

FIG. 3. Measurement set for 60nm device. (a) $\delta R$-$H_{\parallel}$ (black) and $\delta R$-$H_{\perp}$ (gray) loops at -5mV bias. (b) P-state $N$-$I_e$ loops at $H_{\parallel} \approx +1.2$ kOe (red), and AP-state $N$-$I_e$ loops at $H_{\parallel} \approx -0.45$ kOe ( blue); FL critical currents to determine $\Gamma_{FL}$ (via (6)) enclosed by oval. (c) rms PSD$(f, I_e)$ (normalized to 1 mA) with $I_e$ as indicated by color. Thin black curves are least-squares fits via (7), fitted values for $\alpha_{FL}$, $\alpha_{RL}$ listed on top of graph. Measured $r_{bias}$ and applied field $H$ listed inside graph. Field strength and direction (see Fig. 9) adjusted to achieve "magic-angle". ±1.5 mA spectra shown, but not fit.

As shown previously,[4] narrow-band "low"-frequency $N$-$I_e$ measurements ( $N \equiv \sqrt{PSD(f = 100\,\text{MHz})}$ , 1MHz bandwidth) can reveal spin-torque criticality as the *very rapid onset* of excess (1/*f*-like) noise when $|I_e|$ exceeds $|I_e^{crit}|$. $N$-$I_e$ loops are measured with $I_e$ sourced from a continuous sawtooth generator (2-Hz) which also triggers 1/2 sec sweeps of an Agilent-E4440 spectrum analyzer (in zero-span, averaging mode) for ≈50 cycles. With high sweep repeatability and virtually no $I_e$ -hysteresis, this averaging is sufficient so that after (quadratically) subtracting the mean $N(I_e \approx 0) \approx 1\,\text{nV}/\sqrt{\text{Hz}}$ electronics noise, the resultant $N$-$I_e$ loops (Fig. 3b) indicate stochastic uncertainty $\ll 0.1\,\text{nV}/\sqrt{\text{Hz}}$.

With $\cos\theta = \pm 1$, it readily follows from (3c) and (4) that

$$\Gamma = -\frac{I_e^{crit}(\theta = \pi) \equiv I_{AP}^{crit}}{I_e^{crit}(\theta = 0) \equiv I_{P}^{crit}} \qquad (6)$$

Hence, to estimate $\Gamma$, $N$-$I_e$ are measured with applied fields $H_{\parallel} \approx -0.45, +1.2\,\text{kOe}$ (Fig. 3b),which more than sufficient to align $\hat{\boldsymbol{m}}_{FL}$ antiparallel (AP), or parallel (P) to $\hat{\boldsymbol{m}}_{RL}$ ,respectively (see Fig. 3a), thereby reducing possible sensitivity to Oersted field and/or thermal effects. (Reducing $|H_{\parallel}|$ by ~200-300 Oe did not significantly change either $N$-$I_e$ curve.) With $I_e > 0$ denoting *electron* flow from RL to FL, it is readily found from (3) that $I_{AP\text{-}FL}^{crit} > 0$ and $I_{P\text{-}FL}^{crit} < 0$ for the FL. By symmetry, it must follow that $I_{AP\text{-}RL}^{crit} < 0$ and $I_{P\text{-}RL}^{crit} > 0$ for spin-torque induced instability of the RL. This sign convention readily identifies these four critical points by inspection of the $N$-$I_e$ data. To account for possible small (thermal) spread in critical onset, specific values for the $I_e^{crit}$ (excluding $I_{P\text{-}RL}^{crit}$) are defined by where the $N$-$I_e$ curves cross the $0.2\,\text{nV}/\sqrt{\text{Hz}}$ line, which is easily distinguished from the $\sim 0.05\,\text{nV}/\sqrt{\text{Hz}}/\text{mA}$ residual magnetic/thermal background. $I_{P\text{-}RL}^{crit}$ is estimated in Fig. 3b (and repeatedly in Figs. 4-7) to be ≈ +4.5 mA. Arbitrariness in the value of $I_e^{crit}$ from using the $0.2\,\text{nV}/\sqrt{\text{Hz}}$ criterion is thought to only be of minor significance for $I_{AP\text{-}RL}^{crit}$, due to the rounded shape of the AP $N$-$I_e$ curves near this particular critical point, which may in part explain why $\Gamma_{RL/Cu}$ estimated from $N$-$I_e$ is found to be systematically somewhat larger than $\Gamma_{Cu/FL}$.

However, the key results here are the 0.1-18 GHz broad-band (rms) PSD$(f; I_e)$ spectra (Fig. 3c). They are measured at discrete dc bias currents with the same Miteq preamp (and in-series bias-T) used for the $N$-$I_e$ data, the latter being *insitu* gain-calibrated vs. frequency (with ≈50Ω preamp input impedance and additionally compensating the present ≈0.7 pF device capacitance) to yield quantitatively absolute values for these $\mathrm{PSD}(f;I_e)$ (each averaged over ~100 sweeps, with $\mathrm{PSD}(f;I_e=0)$ subtracted post-process). To confirm the real existence of an effective "magic-angle", the applied field $H$ was carefully adjusted (by repeated trial and error) in both amplitude and direction to eliminate as much as possible any real-time observed dependence of the raw $\mathrm{PSD}(f;I_e)$ near the FL FMR peak (~ 6 GHz) on the *polarity* as well as amplitude of $I_e$ over a sufficient range. This procedure was somewhat tedious and delicate, and initial attempts using a nominally transverse field $H_\perp$ were empirically found inferior to additionally adjusting the direction of the field, here rotated somewhat toward the pinning direction for the RL. Using a mechanically-positioned permanent magnet as a field source, this field rotation was only crudely estimated at the time to be ~20-30$^o$ (see also Sec. IV). With both $H$ and bias-point $\theta_{\mathrm{bias}}$ "optimized" as such, an $I_e$-series of $\mathrm{PSD}(f;I_e)$ were measured, after which the bias-resistance $R_{\mathrm{bias}}$, and finally $R_{\mathrm{min}}$ and $R_{\mathrm{max}}$ were measured at a common (low) bias of –10 mV to determine $r_{\mathrm{bias}}$ (as in (4)).

The key feature of the rms $\mathrm{PSD}(f;I_e)$ in Fig. 3c is that these measured spectra (excluding $I_e=+1.5\,\mathrm{mA}$) appear essentially independent of both the polarity and magnitude of $I_e$ (after 1mA-normalization), defining a "universal" spectrum curve over the entire 18GHz bandwidth, including the *unexpectedly wide, low amplitude* RL-FMR peak near 14 GHz (more on this below). Because of the relatively large $\sqrt{\mathrm{PSD}(f;I_e=0)} \sim 1\,\mathrm{nV}/\sqrt{\mathrm{Hz}}$ background, these RL peaks were not well discernible during raw spectrum measurements, and were practically revealed only after electronics background noise subtraction. As suggested in Fig. 3c, eventual breakdown of the magic-angle condition was generally found to first occur from spin-torque instability of the FL at larger *positive* $I_e$.

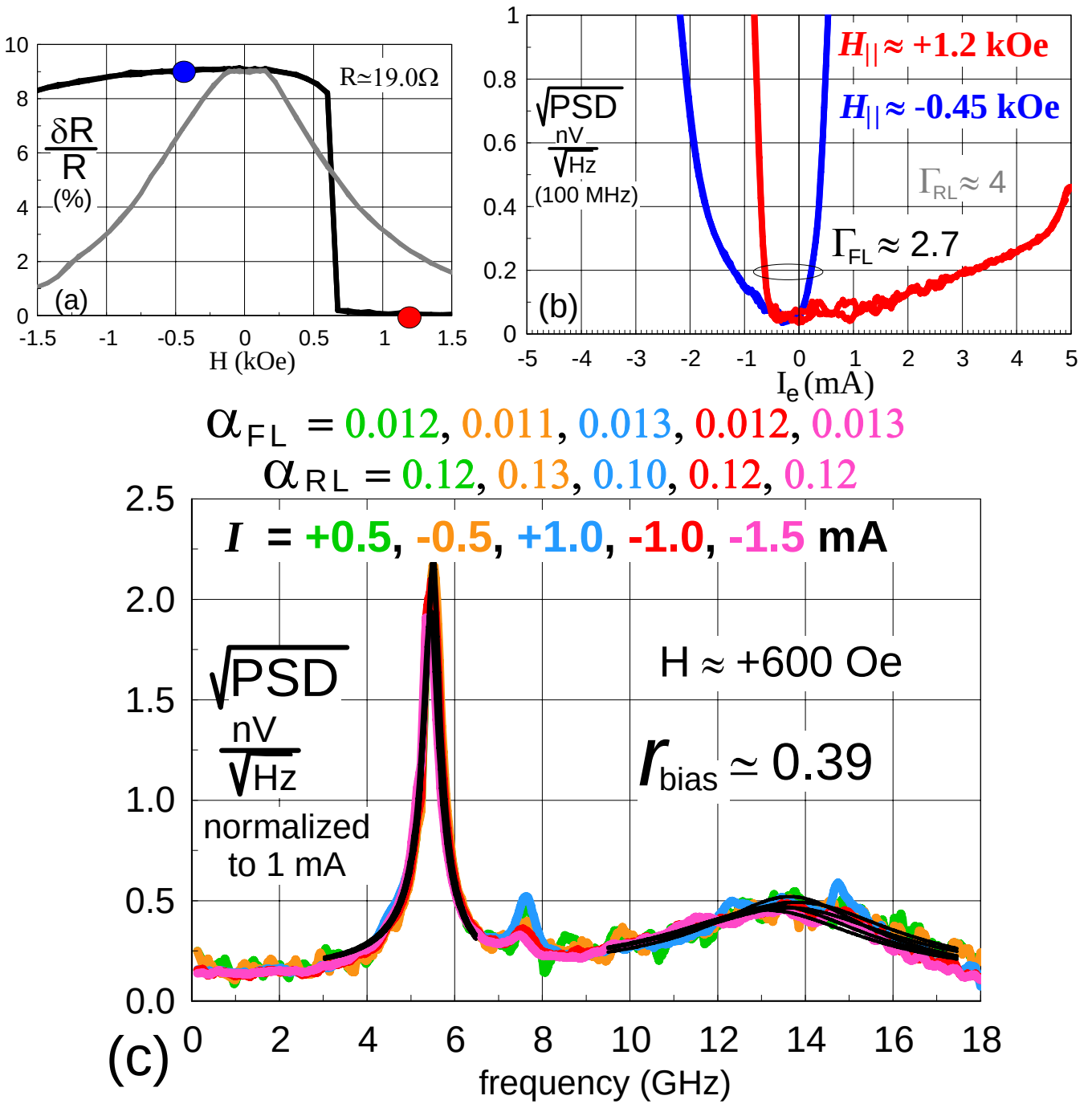

FIG. 4. Analogous measurement set for a different (but nominally identical) 60nm device. as that shown in Fig. 3.

The spectra Fig. 4 shows the equivalent set of measurements on a physically different (though nominally identical) 60-nm device. They are found to be remarkably alike in all properties to those of Fig. 3, providing additional confirmation that the "magic-angle" method can work on real nanoscale structures *to directly obtain the intrinsic* $S_\theta(f;I_e=0)$ *in the absence of of spin-torque effects*. This appears further confirmed by the close agreement of measured $(r_b,\Gamma_{\mathrm{Cu/FL}})$ pairs (from data of Figs. 3,4, and 6) and the macrospin model predictions described in Fig. 2.

To obtain values for linewidth $\Delta\omega$ and then damping $\alpha$ from the measured $\mathrm{PSD}(f;I_e)$, regions of spectra several-GHz wide, surrounding the FL and RL FMR peaks are each nonlinear least-sqaures fitted to the functional form for $S_\theta(f;I_e=0)$ in (5). In particular, the fitting function is taken to be

$$S_V(f=\tfrac{\omega}{2\pi}) = S_0\,\frac{\omega_0^2\,[\omega_0^2+(H'_{y'y'}/H'_{z'z'})\,\omega^2]}{(\omega^2-\omega_0^2)^2+(\omega\,\Delta\omega)^2},$$

$$\text{with } \omega_0=\gamma\sqrt{H'_{y'y'}H'_{z'z'}},\ \Delta\omega=\gamma\,\alpha_{\mathrm{fit}}\,(H'_{z'z'}+H'_{y'y'}) \qquad (7)$$

$$\text{and } H'_{y'y'}\rightarrow[(\omega_{\mathrm{peak}}/\gamma)^2+(\alpha_{\mathrm{fit}}H'_{z'z'})^2/2]/H'_{z'z'}$$

Three fitting parameters are used: $S_0=S_V(f=0)$, $\alpha_{\mathrm{fit}}$, and $\omega_{\mathrm{peak}}$, the latter being already well defined by the data itself. The substitution for $H'_{y'y'}$ is accurate to order $\alpha^2$, leaving $H'_{z'z'}$ as yet unknown. With $H'_{z'z'} \gg H'_{y'y'}$ dominated by out-of-plane demagnetizing fields, $S_V(f)$ depends mostly on the product $\alpha_{\mathrm{fit}}H'_{z'z'}$. For simplicity, fixed values $H'^{\mathrm{FL}}_{z'z'}=8\,\mathrm{kOe}$ and $H'^{\mathrm{RL}}_{z'z'}=10\,\mathrm{kOe}$, were used here, based on macrospin calculations that approximately account for device geometry and net $M_s t$ product for FL and RL films. The fitted $\mathrm{PSD}(f;I_e)$ curves, and the values obtained for $\alpha^{\mathrm{FL}}_{\mathrm{fit}}$ and $\alpha^{\mathrm{RL}}_{\mathrm{fit}}$ are also included in Figs. 3c and 4c. These values are notably *independent* of (or show no significant trend with) $I_e$.

Although the $\alpha^{\mathrm{FL}}_{\mathrm{fit}}\approx 0.01$ repeatedly found from these data is a quite typical magnitude for Gilbert damping in CoFe alloys, the *extremely large*, 10× *greater value of* $\alpha^{\mathrm{RL}}_{\mathrm{fit}}\approx 0.1$ is quite noteworthy, since the RL and FL are not too dissimilar in thickness and composition. Although the small amplitude of the RL-FMR peaks in Figs. 3-4 (everywhere below the raw $1\,\mathrm{nV}/\sqrt{\mathrm{Hz}}$ electronics noise), may suggest a basic unreliability in this fitted value for $\alpha^{\mathrm{RL}}_{\mathrm{fit}}$, this concern is seemingly dismissed by the data of Fig. 5. Measured on a third (nominally identical) device, an alternative "extrapolation-method" was used, in which the applied field was purposefully reduced in magnitude (and more transversely aligned than for magic-angle measurements) to increase $r_{\text{bias}}$ and thus align $\hat{\boldsymbol{m}}_{\text{FL}}$ to be more antiparallel to $\hat{\boldsymbol{m}}_{\text{RL}}$. As a result, spin-torque effects at larger negative-$I_e$ will decrease $\Delta\omega$ and concomitantly *enhance* RL-FMR peak amplitude (and visa-versa for the FL), bringing this part of the measured spectrum above the raw electronics noise background.

Using the same fitting function from (7), it is now necessary to extrapolate the $\alpha_{\text{fit}}^{\text{RL}}(I_e)$ to $I_e \to 0$ (Fig. 5d) in order to obtain the intrinsic damping. This method works well in the case of the RL since $d\alpha_{\text{fit}}^{\text{RL}}/d|I_e| < 0$ and the extrapolated $I_e = 0$ intercept value of $\alpha_{\text{RL}}$ is necessarily *larger* than the measured $\alpha_{\text{fit}}^{\text{RL}}(I_e)$, and hence will be (proportionately) less sensitive to uncertainty in the estimated extrapolation slope. As can be seen from Fig. 5d, *the extrapolated values for intrinsic $\alpha_{\text{RL}}$ are virtually identical to those obtained from the data of Figs. 3,4.* The extrapolated $\alpha_{\text{FL}}$ is also quite consistent as well. The extrapolation data also confirm the expectation (noted earlier following (5)) that linewidth $\Delta\omega$ will vary *linearly* with $I_e$.

Comparing with Figs. 3c,4c, the spectra in Fig. 5c illustrate the profound effect of spin-torque on altering the linewidth and peak-height of both FL and RL FMR peaks even if the system is only moderately misaligned from the magic-angle condition. By contrast, for other frequencies (where the $\Delta\omega$ term in the denominator of (5) is unimportant), the 1mA-normalized spectra are *independent of* $I_e$. Being consistent with (5), this appears to verify that this 2$^{\text{nd}}$ form of fluctuation-dissipation theorem remains valid despite that the system of (1) is not in thermal equilibrium[10] at nonzero $I_e$. (Alternatively stated, spin-torques lead to an *asymmetric* $\overleftrightarrow{H}$, but do not alter the damping tensor $\overleftrightarrow{D}$ in (1)). The $\alpha$-proportionality in the prefactor of $S_\theta(f)$ in (5) relatedly shows that the effect of spin-torque on $\Delta\omega$ is *not* equivalent to additional damping (positive or negative) as may be commonly misconstrued. It further indicates that Oersted-field effects, or other $I_e$-dependent terms in $\overleftrightarrow{H}$ not contributing to $\Delta\omega$, are insignificant in this experiment.

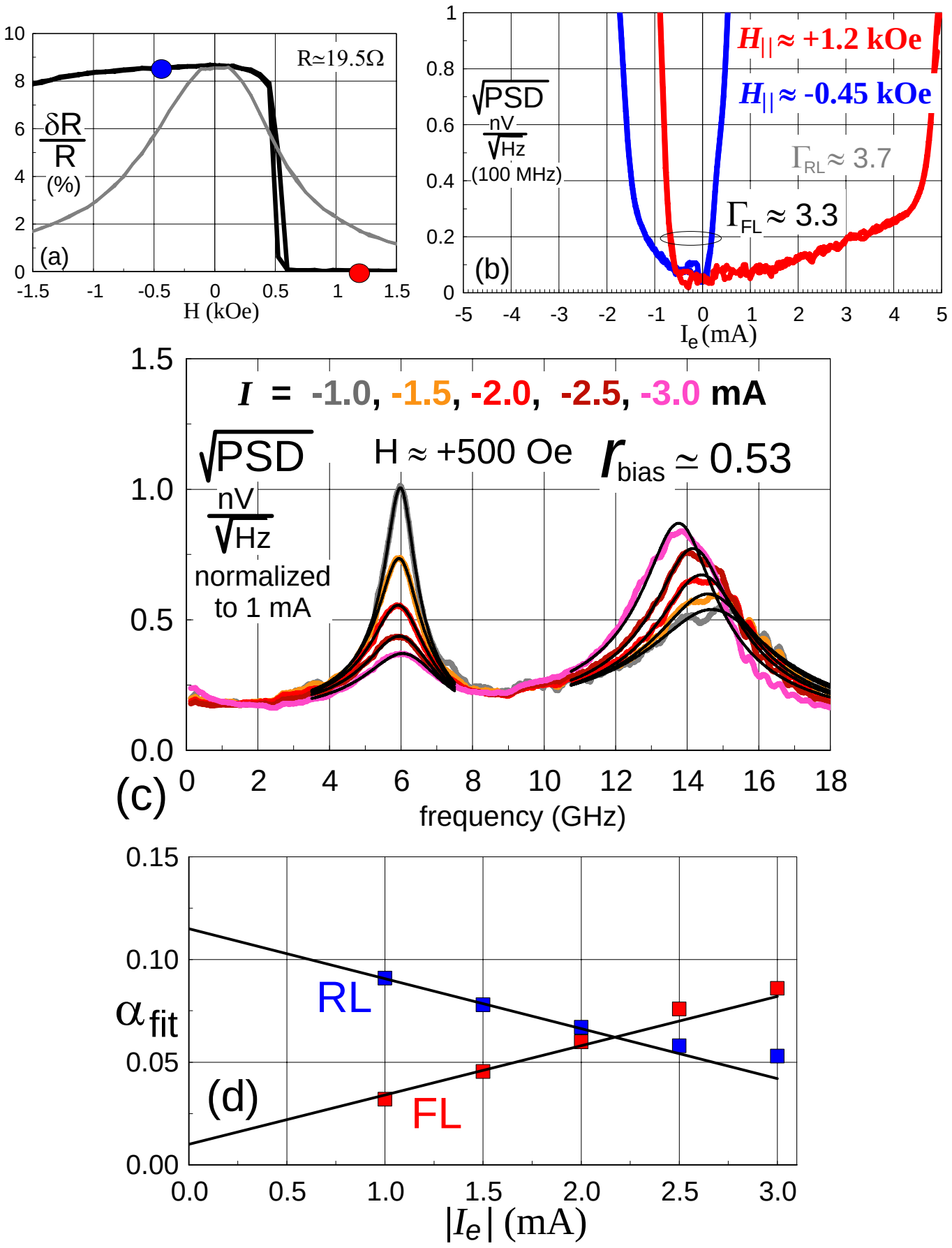

FIG. 5. Measurement set for a different (but nominally identical) 60nm device as that shown in Figs. 3-4. (c) rms spectra (with least-sqaures fits) measured at *larger* $r_{\text{bias}}$ and $\theta_{\text{bias}} > \theta_{\text{magic}}$. (d) $I_e$-*dependent* values of $\alpha_{\text{fit}}(I_e)$ for FL (red) and RL (blue), with suggested $I_e \to 0$ extrapolation lines.

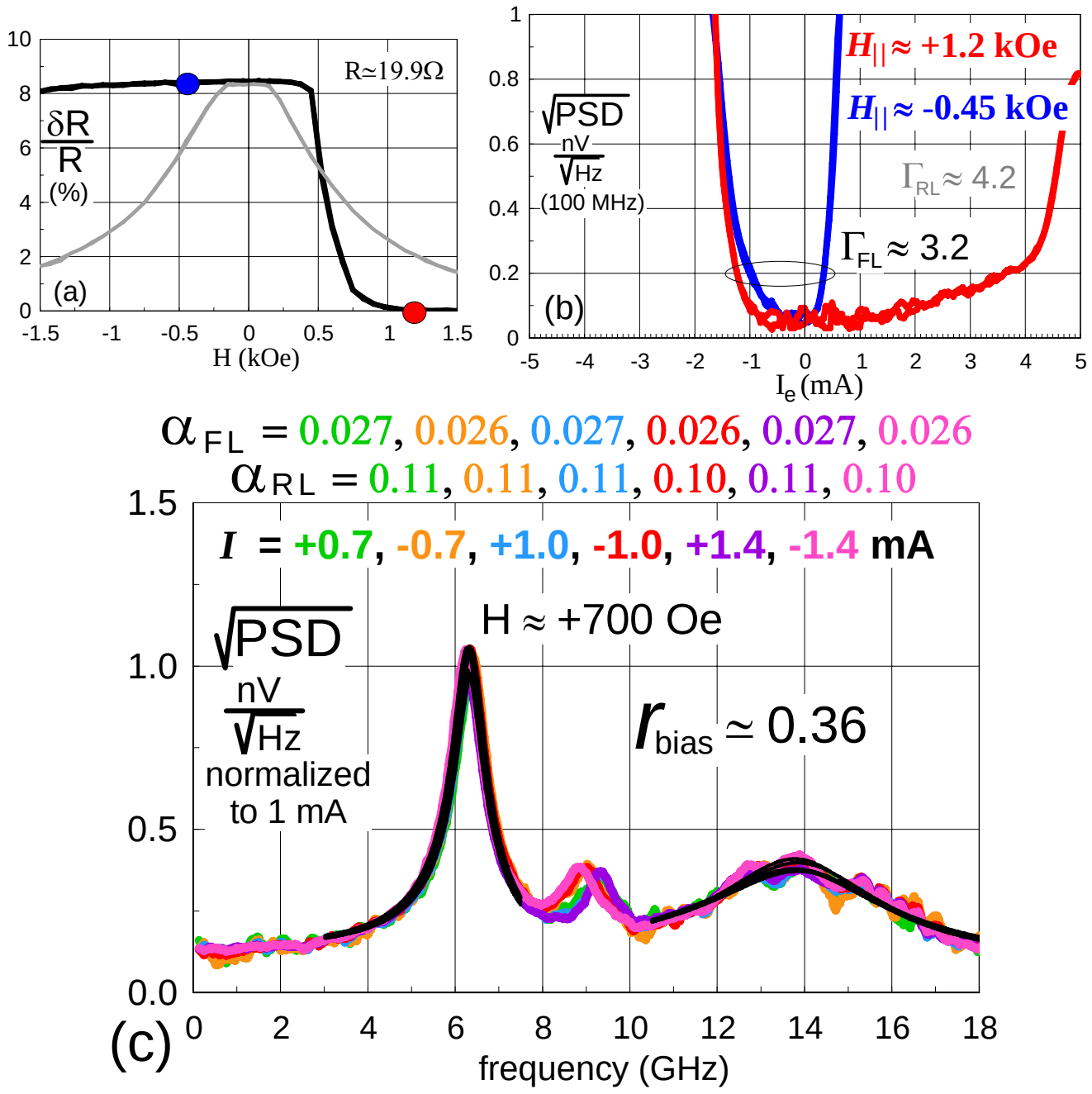

FIG. 6. Analogous measurement set as in Figs.3-4, for (an otherwise identical) device with a 10A Dy cap layer in direct contact with the FL

Analogous to Figs. 4,5, the data of Figs. 6,7 are measured on CPP-GMR-SV stacks differing only by an additional 1-nm thick Dy cap layer deposited directly on top of the FL. The use of Dy in this context (presumed spin-pumping from FL to Dy, but possibly including Dy intermixing near the FL/Dy interface[11]) was found in previous work[12] to result in an ~3× increase in FL-damping, then inferred from the ~3× increase in measured $|I_{\text{FL}}^{\text{crit}}|$. Here, a more direct measure from the FL FMR linewidth indicates a roughly similar, ≈2.3× increase in $\alpha_{\text{FL}}$ (now using somewhat thicker FL films). This ratio is closely consistent with that inferred from $|I_{\text{FL}}^{\text{crit}}|$ data measured in this experiment over a population of devices (see Table 1). Notably, the values found for $\alpha_{\text{RL}}$ remain virtually the same as before.

Finally, Fig.8 shows results for a "synthetic-ferrimagnet" (SF) free-layer of the form FL1/Ru(8A)/FL2. The Ru spacer provides an interfacial *antiferromagnetic* coupling of $\cong 1.0\,\mathrm{erg/cm^2}$. Here, FL1 has a thicker CoFeGe layer than used for prior FL films, and FL2 is a relatively thin CoFe layer chosen so that $(M_s t)_{\mathrm{FL1}} - (M_s t)_{\mathrm{FL2}} \cong (M_s t)_{\mathrm{FL}} \approx 0.64\ \mathrm{erg/cm^2}$. Although having similar static *M-H* or *R-H* characteristics to that of the simple FL (of similar *net* $M_s t$ product) used in earlier measurements, the transport of the SF-FL in regard to spin-torque effects in particular is fundamentally distinct. The basic physics of this phenomenon was described in detail previously.[13] In summary, a spin-torque induced quasi-coresonance between the two natural oscillation modes of the FL1/FL2 couple in the case of *negative* $I_e$ *and* $\hat{\boldsymbol{m}}_{\mathrm{FL1}} \cdot \hat{\boldsymbol{m}}_{\mathrm{RL}} > 0$, can act to transfer energy out of the mode that is destabilized by spin-torque, thereby delaying the onset of criticality and substantially increasing $|I^{\mathrm{crit}}_{\text{P-FL}}|$. Indeed, the side-by-side comparison of $N$-$I_e$ loops provided in Fig. 8b indicate a *nearly 5× increase* in $|I^{\mathrm{crit}}_{\text{P-FL}}|$, despite that $|I^{\mathrm{crit}}_{\text{AP-FL}}|$ remains virtually unchanged.

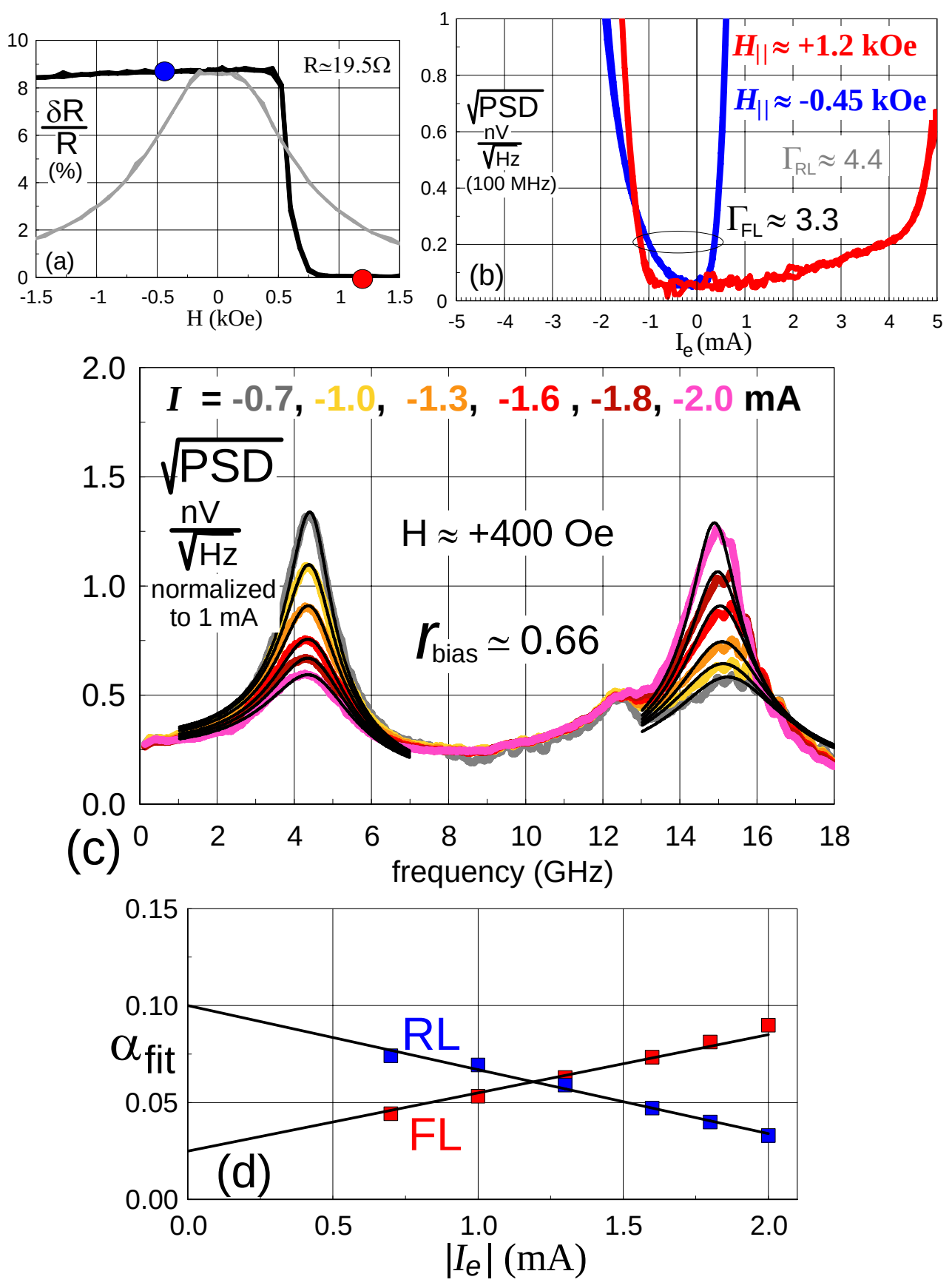

FIG. 7. Analogous measurement set as in Fig. 5 for a different (but nominally identical) device as that in Fig. 6 with a 10A Dy cap layer..

| stack | $-I^{\mathrm{crit}}_{\mathrm{FL}}R$ (mV) | $\alpha_{\mathrm{FL}}$ / $\alpha_{\mathrm{RL}}$ |
|---|---|---|
| Control | 10.4±0.1 | 0.011 / 0.11 |
| Dy cap | 24.5±0.5 | 0.026 / 0.11 |
| SF-FL | 44.9±2.0 | 0.011 / 0.12 |

Table. 1. Summary of critical voltages (measured over ≈ 8 devices each) and damping parameter values α for the present experiment. Estimated statistical uncertainty in the α-values is ~10%.

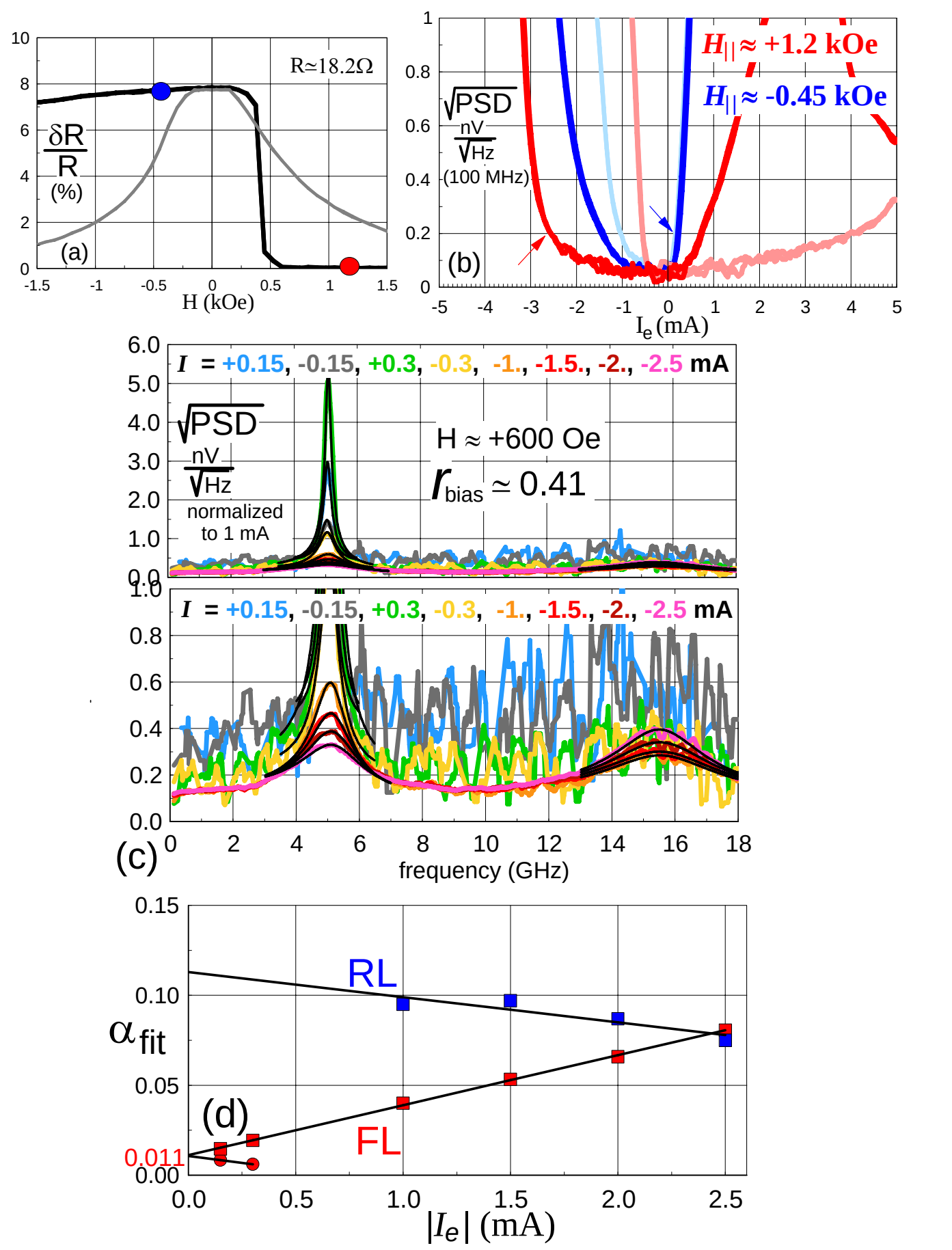

FIG. 8. Analogous measurement set as in Figs.5, for (an otherwise identical) device with a synthetic-ferrimagnet FL (SF-FL) as described in text. (b) includes for comparison $N$-$I_e$ loops (in lighter color) from Fig. 3b; arrows show SF-FL $I_{\mathrm{crit}}$ for P-state (red) and AP-state (blue). (c) spectral data and fits are repeatedly shown (for clarity) using two different ordinate scales.

For the SF-FL devices, attempts at finding the magic-angle under similar measurement conditions as used for Figs. 3c,4c, and 6c were not successful, and so the extrapolation method at similar $r_{\mathrm{bias}} \approx 0.4$ was used instead. To improve accuracy for extrapolated-$\alpha_{\mathrm{FL}}$, the data of Fig. 8c include measurements for $|I_e| \le 0.3\,\mathrm{mA}$ (so that $I_e < I^{\mathrm{crit}}_{\mathrm{FL}}$) for which electronics noise overwhelms the signal from the RF FMR peaks. Showing excellent linearity of $\alpha^{\mathrm{FL}}_{\mathrm{fit}}$ vs. $I_e$ over a wide $I_e$-range, the extrapolated intrinsic $\alpha_{\mathrm{FL}} \approx 0.01$ is, as expected, unchanged from before. The same is true for the extrapolated $\alpha_{\mathrm{RL}}$ as well.

Table 1 summarizes the mean critical *voltages* $-RI^{\mathrm{crit}}_{\text{P-FL}}$ (less sensitive to lithographic variations in actual device area) from a larger set of PSD-$I_e$ measurements. The ≈2.3× increase in $|RI^{\mathrm{crit}}_{\text{P-FL}}|$ with the use of the Dy-cap is in good agreement with that of the ratio of measured $\alpha_{\mathrm{FL}}$.

## IV. MICROMAGNETIC MODELLING

For more quantitative comparison with experiment than afforded by the 1-macrospin model of Sec. II, a 2-macrospin model equally treating both $\hat{\boldsymbol{m}}_{\mathrm{RL}}$ and $\hat{\boldsymbol{m}}_{\mathrm{FL}}$ is now considered here as a simpler, special case of a more general micromagnetic model to be discussed below. The values $M_s^{\mathrm{FL}} = 950\,\mathrm{emu/cc}$, $t_{\mathrm{FL}} = 7\,\mathrm{nm}$, $M_s^{\mathrm{RL}} = 1250\,\mathrm{emu/cc}$, and $t_{\mathrm{RL}} = 5\,\mathrm{nm}$ will be used as simplified, combined representations (of similar thickness and $M_s t$) to the actual CoFe/CoFeGe multilayer films used for the RL and FL. The magnetic films are geometrically modeled as 60 nm squares which (in the macrospin approximation) have zero shape anisotropy (like circles), but allow analytical calculation of all magnetostatic interactions. The effect of IrMn exchange pinning on the RL is simply included as a *uniform* field $\boldsymbol{H}_{\mathrm{pin}} = [J_{\mathrm{pin}}/(M_s t)_{\mathrm{RL}}]\hat{\boldsymbol{x}}$ with measured $J_{\mathrm{pin}} \cong 0.75\,\mathrm{erg/cm}^2$. Firstly, Fig. 9b shows simulated $r_{\mathrm{bias}}$-$H_{\parallel}$ and $r_{\mathrm{bias}}$-$H_{\perp}$ curves computed assuming $\Gamma = 3.2$, roughly the mean value found from the $I_{\mathrm{FL}}^{\mathrm{crit}}$ data of Sec. III. The agreement with the shape of the measured $R$-$H$ is very good (e.g., Figs. 6,7 in particular), which reflects how remarkably closely these actual devices resemble idealized (macrospin) behavior.

Next, Fig. 9d shows simulated PSD curves $S_V(f)$ computed (see Appendix) *in the absence of spin-torque* (i.e., $H_{\mathrm{ST}} = 0$), but otherwise assuming typical experimental values $R$=19Ω, $\Delta R/R$=9%, and $T$=300K, as well as $\alpha_{\mathrm{FL}} = 0.01$ and $\alpha_{\mathrm{RL}} = 0.1$, so to be compared with the magic-angle spectra of Figs. 3,4. Since (as stated in Sec. III) the experimental field angle was not accurately known, the field angle $\phi_H$ was varied systematically for the simulations, and in each case the field-magnitude $H$ was iterated until $r_{\mathrm{bias}} \cong 0.37$, approximately matching the mean measured value. In terms of *both* absolute values *and* the ratio of FL to RL FMR peak amplitudes, the location of $f_{\mathrm{peak}}$ (particularly for the FL), and the magnitude of $H$ (on average 650-700 Oe from the three magic-angle data in Sec. III), the best match with experiment clearly occurs with $30^{\circ} \le \phi_H \le 40^{\circ}$. The agreement, both qualitatively and quantitatively, is again remarkable given the simplicity of the 2-macrospin model.

Finally, results from a discretized micromagnetic model are shown in Fig. 10. Based on Fig.9, the value $\theta_H = 35^{\circ}$ was fixed, and $H$ = 685 Oe was determined by iteration until $r_{\mathrm{bias}} \cong 0.37$. The equilibrium bias-point magnetization distribution is shown in Fig. 10a for this 416 cell model. Estimated values for exchange stiffness, $A_{\text{FL}} = 1.4\,\mu\text{erg/cm}$ and $A_{\text{RL}} = 2\,\mu\text{erg/cm}$ were assumed. The simulated spectra in Fig. 10b are shown *one eigenmode at a time* (see Appendix), for the 7 eigenmodes with predicted FMR frequencies below 20 GHz (the $8^{\text{th}}$ mode is at 22.9 GHz). The $1^{\text{st}}$, $2^{\text{nd}}$, $5^{\text{th}}$, and $7^{\text{th}}$ modes involve mostly FL motion, the nearly degenerate $3^{\text{rd}}$ and $4^{\text{th}}$ modes (and the $6^{\text{th}}$) mostly that of the RL. (The amplitudes of $S_V(f)$ from the $6^{\text{th}}$ or $7^{\text{th}}$ mode are negligible.). For illustration purposes only, Fig. 10b assumed identical damping $\alpha_{\text{FL}} = \alpha_{\text{FL}} = 0.01$ in each film.

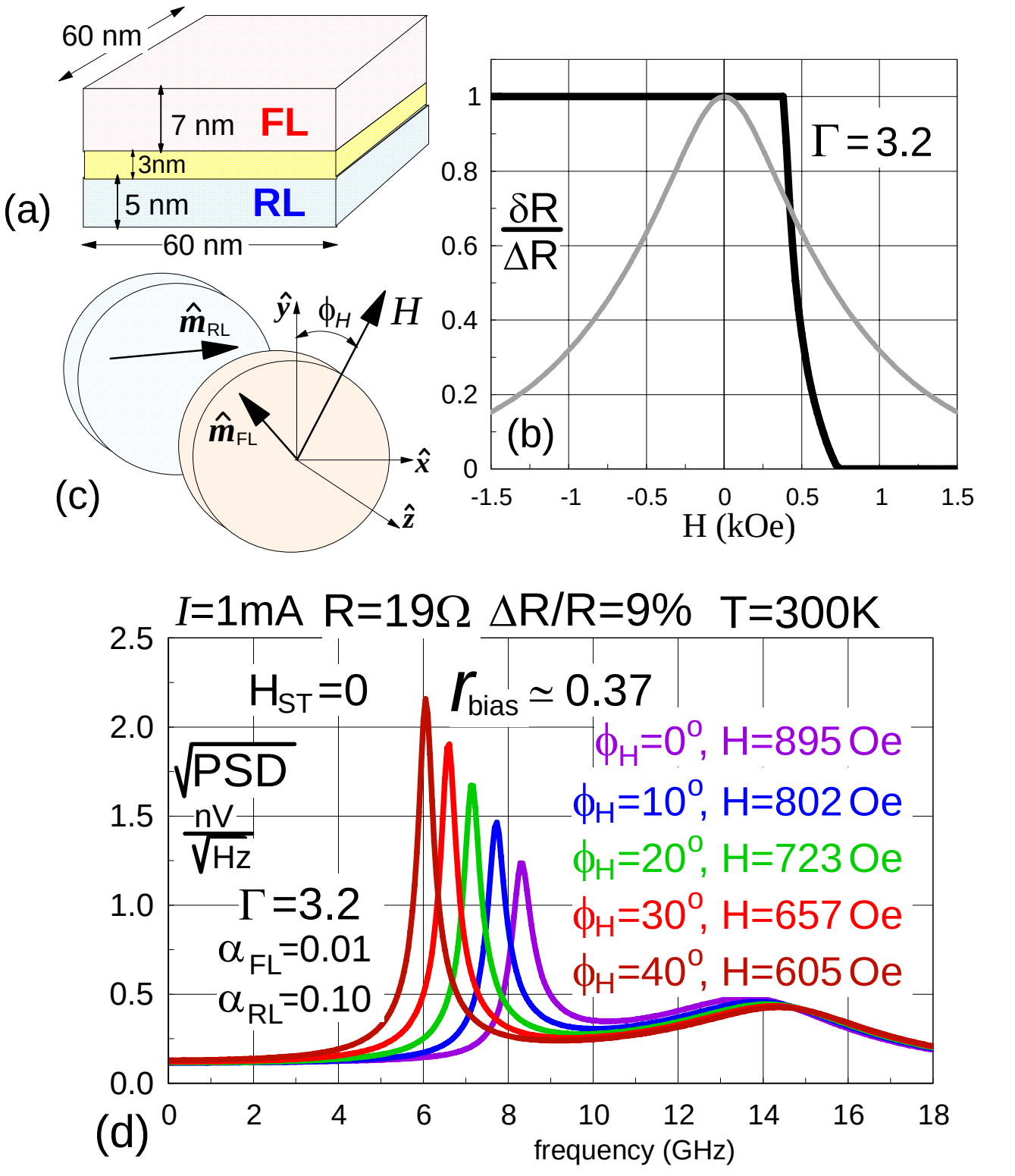

FIG. 9. Two-macrospin model results. (a) cartoon of model geometry. (b) simulated $\delta R$-$H$ loops analogous to data of Figs 3-8c. (c) cartoon defining vector orientations (RL exchange pinned along +$\boldsymbol{x}$ direction). (d) simulated rms PSD assuming parameter values indicated, with variable $|H|$ to maintain a fixed $r_{\mathrm{bias}}$ at each $\phi_H$ (as indicated by color).

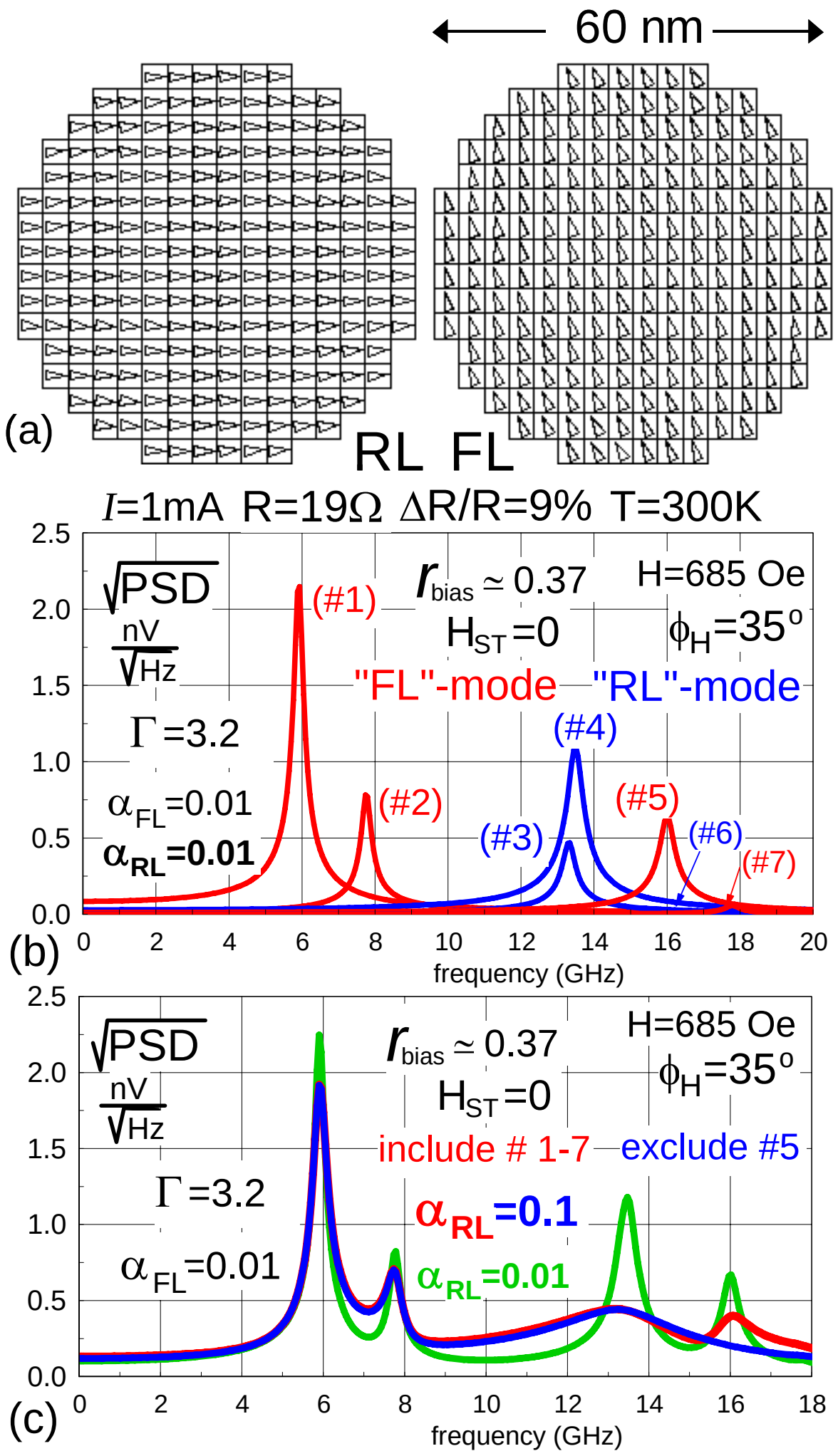

FIG. 10. Micromagnetic model results. (a) cell discretizations with arrow-heads showing magnetization orientation when $|H|$=685 Oe and $\phi_H = 35^{\circ}$ (see Fig. 9c). (b) simulated partial rms PSD for first 7 eigenmodes (as labeled) computed *individually* with $\alpha_{\mathrm{FL}} = 0.01$ *and* $\alpha_{\mathrm{RL}} = 0.01$, other parameter values indicated. (c) simulated total rms PSD with $\alpha_{\mathrm{FL}} = 0.01$ and $\alpha_{\mathrm{RL}} = 0.01$ (green) or $\alpha_{\mathrm{RL}} = 0.1$ (red or blue); blue curve excludes contribution from 5th (FL) eigenmode at 16 GHz.

For Fig. 10c, the computation of $S_V(f)$ is more properly computed using either 6 or all 7 eigenmodes simultaneously, which includes damping-induced coupling between the modes. Including higher order modes makes negligible change to $S_V(f < 20\,\text{GHz})$ (but rapidly increases computation time). As was observed earlier, the agreement between simulated and measured spectra in Figs. 3c,4c is good (with $\alpha_{\text{RL}} = 0.1$), and the simulations now include the small, secondary FL-peak near 8 GHz clearly seen in the measured data (including that of Fig. 6c), though it is somewhat more pronounced in the model results. Notably, the computed spectrum near the RL FMR peak more resembles the measurements after removing the 16-GHz $5^{\text{th}}$ mode from the calculation, as this (FL) mode does not appear to be physically present in the Figs. 3c,4c spectra.

While it is perhaps expected that higher order modes in a micromagnetic simulation assuming perfectly homogeneous magnetic films would show deviations from real devices with finite grain-size, edge-roughness/damage, etc., the situation is actually more interesting. Fig. 11 shows measured spectra on yet another device (again, nominally identical to that of Figs. 3-5) in which the experiment was perhaps slightly off from the optimum magic-angle condition, as evidenced by the very small shift in the 6-GHz FL FMR peak position with polarity of $I_e$. More noteworthy, however, is the clear polarity asymmetry and nonlinear-in- $I_e$ peak-amplitude (for $I_e > 0$ in particular) of *both* the 8-GHz secondary FL mode *and* a higher order mode close to 15 GHz. (Both resemble typical spin-torque effect at angles more antiparallel than the magic-angle condition.) This similarity in behavior indicates with near certainty that this 15-GHz mode is also FL-like in origin, and is thus a demonstration of the "missing" $5^{\text{th}}$ mode predicted in Fig. 10. (In hindsight, there is now discernible a small but similar 15-GHz peak in the spectra of Fig. 4c). It is worth remembering that the "magic-angle" argument was based on a simple 1-macrospin model, and so remarkably there appears to be circumstances where this "spin-torque null" actually does apply simultaneously to both the FL and RL, as well as to higher order modes.

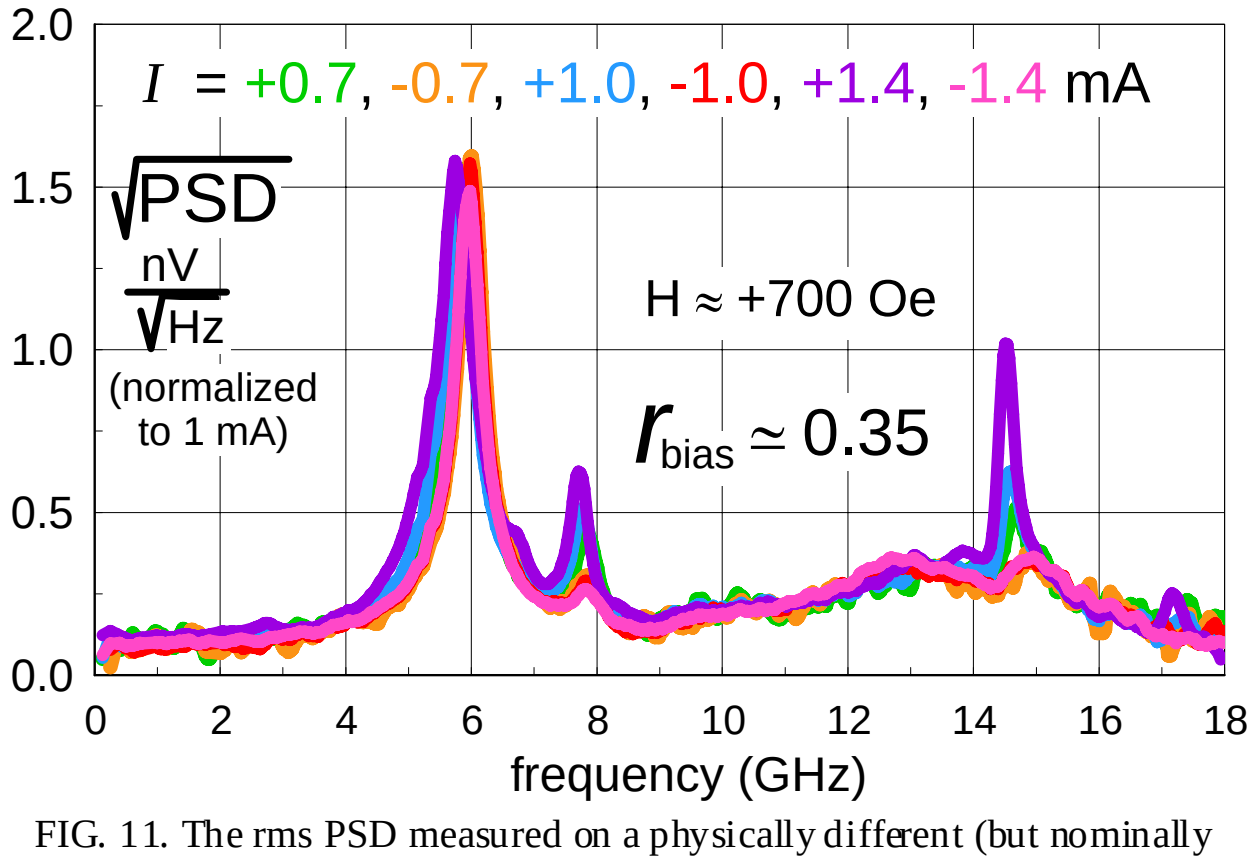

FIG. 11. The rms PSD measured on a physically different (but nominally identical) device as that generating the analogous "magic-angle" spectra shown in Figs. 3c and 4c.

## V. DISCUSSION

In addition to the direct evidence from the measured spectral linewidth in Figs. 3-8, evidence for large Gilbert damping $\alpha_{\text{RL}} >> \alpha_{\text{FL}}$ for the RL is also seen in the $I_e^{\text{crit}}$ data. As ratios $I_{\text{P-RL}}^{\text{crit}} / I_{\text{P-FL}}^{\text{crit}}$ and $I_{\text{AP-RL}}^{\text{crit}} / I_{\text{AP-FL}}^{\text{crit}}$ are (from Figs. 3-5 data) both roughly ~7, this conclusion is semi-quantitatively consistent with the basic scaling (from (3c)) that $I_e^{\text{crit}} \propto \alpha$. This, as well as the substantial, 2-3× variation of $\alpha_{\text{fit}}^{\text{RL}}$ with $I_e$ in Figs. 5d, and 7d, appears to rather conclusively (and expectedly) confirm that inhomogeneous broadening is not a factor in the large linewidth-inferred values of $\alpha_{\text{RL}}$ found in these *nanoscale* spin-valves.

Large increases in effective damping of "bulk" samples of ferromagnetic (FM) films in contact with antiferromagnet (AF) exchange pinning layers has been reported previously.[14-16] The excess damping was generally attributed to two-magnon scattering processes[17] arising from an inhomogeneous AF/FM interface. However, the two-magnon description applies to the case where the uniform, $(\boldsymbol{k} = 0\,, \omega \equiv \omega_0)$ mode is pumped by a external rf source to a high excitation ( magnon) level, which then transfers energy via two-magnon scattering into a *large* (quasi-continuum) number of *degenerate* $(\boldsymbol{k} \neq 0, \omega_{\boldsymbol{k}} = \omega_0)$ spin-wave modes, all with low (thermal) excitation levels and mutually coupled by the same two-magnon process. In this circumstance, the probability of energy transfer back to the uniform mode (just one among the degenerate continuum) is negligible, and the resultant one-way flow of energy out of the uniform mode resembles that of intrinsic damping to the lattice. By contrast, for the *nanoscale* spin-valve device, the relevant eigenmodes (Fig. 10) are discrete and generally *nondegenerate.* in frequency. Even for a coincidental case of a quasi-degenerate pair of modes (e.g., RL modes #3 and #4 in Fig. 10), both modes are equally excited to thermal equilibrium levels (as are all modes), and have similar intrinsic damping rates to the lattice. Any additional energy transfer via a two-magnon process should flow both ways, making impossible a large (e.g., ~10×) increase in the effective net damping of either mode.

Two alternative hypotheses for large $\alpha_{\text{RL}}$ which are essentially independent of device size are 1) large spin-pumping effect at the IrMn/RL interface, or 2) strong interfacial exchange coupling at the IrMn/RL resulting in non-resonant coupling to high frequency modes in either the RL *and/or or the IrMn film.* However, these two alternatives can be distinguished since the exchange coupling strength can be greatly altered without necessarily changing the spin-pumping effect. In particular, $\alpha_{\text{RL}}$ was very recently measured by conventional FMR methods by Mewes[18] on bulk film samples (grown by us with the same RL films and IrMn annealing procedure as that of the CPP-GMR-SV devices reported herein) of the reduced stack structure: seed/IrMn($t_{\rm AF}$)/Cu($t_{\rm Cu}$)/RL/Cu(30A)/cap. For all four cases described in Table 2, the exchange coupling was deliberately reduced to zero, and the measured $\alpha_{\rm RL} \approx 0.012$ was found to be nearly identical to that found here for the FL of similar CoFeGe composition. However, for the two cases with $t_{\rm AF}$ = 60A, excess damping due to spin-pumping of electrons from RL into IrMn should not have been diminished (e.g., the spin diffusion length in Cu is ~100× greater than $t_{\rm Cu}$≈ 30A). This would appear to rule out the spin-pumping hypothesis.

| sample type | $\alpha_{\rm RL} = \frac{\sqrt{3}}{2}\gamma\,(d\Delta H/d\omega)$ |
|---|---|
| $t_{\rm AF} = t_{\rm Cu} = 0$ | 0.011 |
| $t_{\rm AF}$= 60A, $t_{\rm Cu}$ = 0 (no seed layer for IrMn) | 0.010 |
| $t_{\rm AF}$= 60A, $t_{\rm Cu}$ = 30A | 0.013 |
| $t_{\rm AF} = t_{\rm Cu} = 0$ (out-of-plane FMR) | 0.013 |

Table. 2. Summary of bulk film FMR measurements[18] for reduced film stack structure: seed/IrMn($t_{\rm AF}$)/Cu($t_{\rm Cu}$)/RL/Cu(30A)/cap. Removal of IrMn, or alternatively a lack of proper seed layer and/or use of a sufficiently thick $t_{\rm Cu}$≈30A can each effectively eliminate exchange pinning strength to RL.

The second hypothesis emphasizes the possibility that the energy loss takes place inside the IrMn, from oscillations excited far off resonance by locally strong interfacial exchange coupling to a fluctuating $\hat{\boldsymbol m}_{\rm RL}$. This local interfacial exchange coupling $J_{\rm ex}$ can be much greater than $J_{\rm pin}$, since the latter reflects a surface average over inhomogeneous spin-alignment (grain-to-grain and/or from atomic roughness) within the IrMn sub-lattice that couples to the RL. Further, though such strong but inhomogeneities coupling cannot truly be represented by a uniform $\boldsymbol H_{\rm pin}$ acting on the RL, the similarity between measured and modeled values of ~14 GHz for the "uniform" RL eigenmode has clearly been demonstrated here. Whatever are the natural eigenmodes of the real device, the magic-angle spectrum measurements of Sec. III reflect the thermal excitation of all eigenmodes for which "one-way" intermodal energy transfer should be precluded by the condition of thermal equilibrium and the orthogonality[19] of the modes themselves. Hence, without an additional energy sink *exclusive* of the RL/FL spin-lattice system, the linewidth of all modes should arguably reflect the intrinsic Gilbert damping of the FL or RL films, which the data of Sec. III and Table 2 indicate are roughly equal with $\alpha \sim 0.01$. Inclusion of IrMn as a combined AF/RL system, would potentially provide that extra energy loss channel for the RL modes.

A rough plausibility argument for the latter may be made with a crude AF/FM model in which a 2-sublattice AF film is treated as two ferromagnetic layers (#1 and #2) occupying the *same* physical location. Excluding magnetostatic contributions, the free energy/area for this 3-macrospin system is taken to be

$$(M_s t)_{\rm AF} H_{\rm AF}\, \hat{\boldsymbol m}_1 \cdot \hat{\boldsymbol m}_2 - \tfrac{1}{2}(Kt)_{\rm AF}[(\hat{\boldsymbol m}_1 \cdot \hat{\boldsymbol x})^2 + (\hat{\boldsymbol m}_2 \cdot \hat{\boldsymbol x})^2] - J_{\rm ex}\hat{\boldsymbol m}_2 \cdot \hat{\boldsymbol m}_{\rm FM} + [J_0 - J_{\rm pin}]\,\hat{\boldsymbol m}_{\rm FM} \cdot \hat{\boldsymbol x} \qquad (8)$$

For IrMn with Neel temperature of $T_{\rm N} \approx 700K$, the internal AF exchange field $H_{\rm AF} \sim k_{\rm B}T_N/\mu_{\rm B} \sim 10^7$ Oe.[20] With $t_{\rm AF} = 60$A, AF uniaxial anisotropy is estimated to be $K_{\rm AF} \sim 10^6$ erg/cc.[21] A rough estimate $J_{\rm ex} \sim 8(A/t)_{\rm FM}$ for *strong* interfacial exchange is obtained by equating interface energy $J_{\rm ex}\,\bar\phi^2/2$ to the bulk exchange energy $4A\bar\phi^2/t$ of a hypothetical, small angle Bloch wall $(0 \le \phi \le 2\bar\phi)$ twisting through the FM film thickness. Taking $t \approx 5\,{\rm nm}$ and $A$ ~ $10^{-6}$ erg/cm yields $J_{\rm ex} \sim 15\,{\rm erg/cm^2}$. The value of $J_0 \sim [1/J_{\rm ex} + 1/(Kt)_{\rm AF}]^{-1}$ in the last "field-like" term in (8) is more precisely chosen to maintain a constant eigenfrequency for the FM layer *independent* of $J_{\rm ex}$ or $K_{\rm AF}$, thus accounting for the weaker inhomogeneous coupling averaged over an actual AF/FM interface.

As shown in Fig. 12, this crude model can explain a ~10× increase in the FM linewidth provided $J_{ex} \gtrsim 5\text{-}10\,{\rm erg/cm^2}$ and $\alpha_{\rm AF} \sim 0.05\text{-}0.1$. It is worthily noted[20] that for the 2-sublattice AF, the linewidth $\Delta\omega/\omega_0 \approx \alpha_{\rm AF}\sqrt{2H_{\rm AF}/(H_K \equiv (K/M_s)_{\rm AF})}$ is larger by a factor of $\sqrt{2H_{\rm AF}/H_K} \sim 100$ compared to high order FM spin-wave modes in cases of comparable $\alpha$ and $\omega_0$ (with $\omega_0 \approx \gamma\sqrt{2H_{\rm AF}H_K} \sim 10^{12}$ Hz for the AF). Since the *lossy* part of the "low" frequency susceptibility for FM or AF modes scales with $\Delta\omega$, it is suggested that the IrMn layer can effectively sink energy from the ~14 GHz RL mode despite the ~100× disparity in their respective resonant frequencies. S*ize-independent* damping mechanisms for FM films exchange-coupled to AF layers such as IrMn are worthy of further, detailed study.

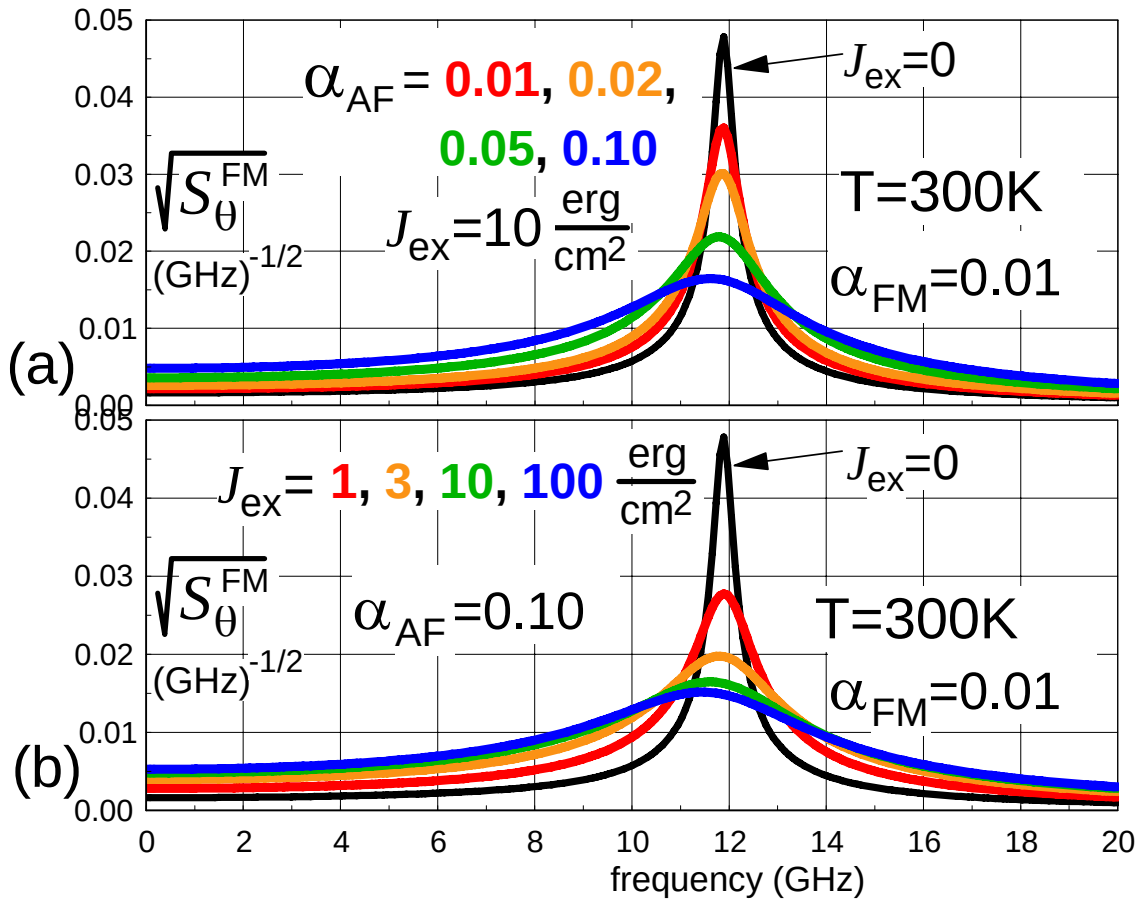

FIG. 12. Simulated rms PSD $S_\theta^{\rm FM}(f)$ for a 3-macrospin model of an AF/FM couple as described via (8) and in the text. The FM film parametrics are the same as used for macrospin RL model in Fig. 9, with $\alpha_{\rm FM}$ =0.01 and $J_{\rm pin}$ = 0.75 erg/cm². (a) varied $\alpha_{\rm AF}$ (denoted by color) with $J_{\rm ex}$=10erg/cm². (b) varied $J_{\rm ex}$ (denoted by color) with $\alpha_{\rm AF}$ =0.1. The black curve in (a) or in (b) corresponds to $J_{\rm ex}$=0. For AF, $M_s$ is taken to be 500 emu/cc.

## ACKNOWLEDGMENTS

The authors wish to acknowledge Jordan Katine for the e-beam lithography used to make all the measured devices, and Stefan Maat for film growth of alternative CPP spin-valve stacks useful for measurements not included here. The authors wish to thank Tim Mewes (and his student Zachary Burell) for making the bulk film FMR measurements on rather short notice. One author (NS) would like to thank Thomas Schrefl for a useful suggestion for micromagnetic modeling of an AF film.

## APPENDIX

As was described in detail elsewhere,[22] the generalization of (1) or (5) from a single macrospin to that for an $N$-cell micromagnetic model takes the form

$$(\vec{\boldsymbol{D}}+\vec{\boldsymbol{G}})\cdot\frac{d\vec{\boldsymbol{m}}'}{dt}+\vec{\boldsymbol{H}}'\cdot\vec{\boldsymbol{m}}'=\vec{\boldsymbol{h}}'(t)$$
$$\vec{\boldsymbol{S}}(\omega)\equiv\frac{2k_BT}{\gamma\Delta m}\vec{\chi}\cdot\vec{\boldsymbol{D}}\cdot\vec{\chi}^{\dagger},\quad \vec{\chi}(\omega)=[\vec{\boldsymbol{H}}'-i\omega(\vec{\boldsymbol{D}}+\vec{\boldsymbol{G}})]^{-1} \qquad \text{(A1)}$$

where $\vec{\boldsymbol{m}}'$ (or $\vec{\boldsymbol{h}}'$) is an $2N\times 1$ column vector built from the $N$ 2D vectors $\boldsymbol{m}'_{j=1\ldots N}$, and $\vec{\boldsymbol{D}},\vec{\boldsymbol{G}},$ and $\vec{\boldsymbol{H}}$ are $2N\times 2N$ matrices formed from the $N\times N$ array of 2D tensors $\vec{D}_{jk}$, $\vec{G}_{jk}$, and $\vec{H}_{jk}$. Here, $\vec{D}_{jk}=\vec{D}\,\delta_{jk}$ and $\vec{G}_{jk}=\vec{G}\,\delta_{jk}$, though $\vec{H}_{jk}$. is *nonlocal* in cell indices $j,k$ due to the magnetostatic interaction.

The PSD $S_Q(f)$ for any scalar quantity $Q(\{\hat{\boldsymbol{m}}_j\})$ is[22]

$$S_Q(f)=2\sum_{j,k=1}^{N}\boldsymbol{d}'_j\cdot\vec{S}_{jk}(\omega)\cdot\boldsymbol{d}'_k,\ \boldsymbol{d}'_j\equiv\frac{\partial\boldsymbol{m}'_j}{\partial\hat{\boldsymbol{m}}_j}\cdot\frac{\partial Q}{\partial\hat{\boldsymbol{m}}_j} \qquad \text{(A2)}$$

The computations for the PSD of Figs.9, 10 took $Q(\vec{\boldsymbol{m}})$ to be

$$Q=\frac{1}{N_i}\sum_{i=1}^{N_i}\frac{I_{\text{bias}}\Delta R\,(1-\hat{\boldsymbol{m}}_i^{\text{RL}}\cdot\hat{\boldsymbol{m}}_i^{\text{FL}})}{\Gamma+1\,+(\Gamma-1)\hat{\boldsymbol{m}}_i^{\text{RL}}\cdot\hat{\boldsymbol{m}}_i^{\text{FL}}}$$

averaged over the $N_i=N/2$ cell pairs at the RL-FL interface.

For a *symmetric* $\vec{\boldsymbol{H}}$ (e.g., $H_{\text{ST}}=0$), the set of eigenvectors $\vec{\boldsymbol{e}}\leftarrow\vec{\boldsymbol{m}}$ of the system (A1) can be defined from the following eigenvalue matrix equation

$$(\vec{\boldsymbol{G}}^{-1}\cdot\vec{\boldsymbol{H}})\cdot\vec{\boldsymbol{e}}_{n=1\ldots 2N}=i\omega_n\,\vec{\boldsymbol{e}}_n \qquad \text{(A3)}$$

The eigenvectors come in $N$ complex conjugate pairs $\vec{\boldsymbol{e}}^{+},\vec{\boldsymbol{e}}^{-}$ with real eigenfrequencies $\pm\omega$. With suitably normalized $\vec{\boldsymbol{e}}_n$, matrices $H_{mn}=\delta_{mn}$ and $G_{mn}=\delta_{mn}/i\omega_n$ are diagonal in the eigenmode basis.[22]. The analogue to (A1) becomes

$$[\chi^{-1}(\omega)]_{mn}=(1-\omega/\omega_n)\,\delta_{mn}-i\omega(D_{mn}\equiv\vec{\boldsymbol{e}}_m^{*}\cdot\vec{\boldsymbol{D}}\cdot\vec{\boldsymbol{e}}_n)$$
$$S_{mn}(\omega)\equiv\frac{2k_BT}{\gamma\Delta m}\sum_{m',n'}\chi_{mm'}(\omega)\,D_{m'n'}\,\chi^{*}_{nn'}(\omega) \qquad \text{(A4)}$$
$$S_Q(f)=2\sum_{m,n}d'^{*}_m\,S_{mn}(\omega)\,d'_n,\ d'_n\equiv\vec{\boldsymbol{e}}_n^{*}\cdot\vec{\boldsymbol{d}}'$$

The utility of eigenmodes for computing PSD, e.g, in the computations of Fig. 10, is that only a *small* fraction (e.g., 7 rather than 416 eigenvector pairs) need be kept in (A4) (with all the rest *simply ignored*) in order to obtain accurate results in practical frequency ranges (e.g., $<20$ GHz). Despite that $D_{mn}$ is (in principle) a full matrix, the reduction in matrix size for the matrix inversion to obtain $\chi(\omega)$ *at each frequency* more than makes up for the cost of computing the $(\omega_n,\vec{\boldsymbol{e}}_n)$ which need be done only once independent of frequency or $\alpha$-values.

[18] T. Mewes, *MINT* Center, U. Alabama (Tuscaloosa). These bulk film FMR measurements were made in-plane, except for the last entry in Table 2 which was measured in the out-of-plane configuration which should exclude two-magnon contributions (Ref. 17). The similarity in $\alpha$ -values suggests two-magnon is unimportant here, which in turn then suggests that RL-FL spin-pumping contributions in the present CPP-GMR nanopillars spin-valves (footnote (5)) that would not be present in the RL-only bulk samples are also small. Further comparisons of present measurements with additional bulk film FMR results are expected to be addressed in a future publication.

[19] A nonzero damping matrix $\overleftrightarrow{D}$ can technically cause a coupling of nondegenerate eigenmodes if $\vec{\boldsymbol{e}}^{*}_{m\neq n} \cdot \overleftrightarrow{\boldsymbol{D}} \cdot \vec{\boldsymbol{e}}_{n} \neq 0$ , though this does not change the argument citing this footnote. This small effect can be seen in Fig. 10c as the slight change in the 6-GHz FL FMR peak height and linewidth when $\alpha_{\mathrm{RL}}$ is grossly varied from 0.01 to 0.1.

[20] *Magnetic Oscillations and Waves*, A. G. Gurevich and G. A. Melkov, CRC Press (Florida, USA) 1996. Chap. 3.

[21] K. O'Grady, L. E. Ferndandez-Outon, G. Vallejo-Fernandez, J. Magn. Magn. Mater. **322**, 883 (2010).

[22] Appendix A of N. Smith, J. Magn. Magn. Mater. **321**, 531 (2009).